\documentclass[conference]{ieeeconf}

\IEEEoverridecommandlockouts
\usepackage{amsmath,amssymb,amsfonts}
\usepackage{algorithmic, algorithm}
\usepackage{graphicx}
\usepackage{textcomp}
\usepackage{xcolor}
\usepackage[hidelinks]{hyperref}
\usepackage{rotating, tabularray, pifont}
\usepackage{subfloat}
\def\BibTeX{{\rm B\kern-.05em{\sc i\kern-.025em b}\kern-.08em
    T\kern-.1667em\lower.7ex\hbox{E}\kern-.125emX}}

\usepackage{subcaption} 

\def\R{\mathbb{R}}

\DeclareMathOperator{\proj}{proj}
\DeclareMathOperator{\prox}{prox}
\DeclareMathOperator*{\argmin}{argmin}
\begin{document}

\title{\LARGE \bf PRIME: Fast Primal-Dual Feedback Optimization for Markets with Application to Optimal Power Flow}

\makeatletter
\newcommand{\newlineauthors}{
  \end{@IEEEauthorhalign}\hfill\mbox{}\par
  \mbox{}\hfill\begin{@IEEEauthorhalign}
}
\makeatother

\author{Nicholas Julian Behr, Mattia Bianchi, Keith Moffat, Saverio Bolognani, and  Florian Dörfler
\thanks{The authors are with the Automatic Control Laboratory, ETH Zürich, Switzerland. This work is supported by ETH Zürich funds and by the Swiss National Science Foundation under the NCCR Automation (grant agreement 51NF40\_225155).}
}

\maketitle

\begin{abstract}
Online Feedback Optimization (OFO) controllers iteratively drive a plant to an optimal operating point that satisfies input and output constraints, relying solely on the input-output sensitivity as model information. 
This paper introduces PRIME (PRoximal Iterative MarkEts), a novel OFO approach based on proximal-point iterations.
Unlike existing OFO solutions, PRIME admits a market-based implementation, where self-interested actors are incentivized to make choices that result in safe and efficient operation, without communicating private costs or constraints.
Furthermore, PRIME can handle non-smooth objective functions, achieve fast convergence rates and rapid constraint satisfaction, and effectively reject measurement noise.
We demonstrate PRIME on an AC optimal power flow problem, obtaining an efficient real-time nonlinear local marginal pricing scheme.
\end{abstract}

\section{Introduction}
Regulating a physical plant to a cost-optimal operating point is a fundamental goal in control theory \cite{control_review, constrained_KKT}. 
The classic solution approach involves pre-computing a reference signal, which an online controller then tracks in real-time. Feedforward reference optimization is, however, not robust to uncertainty in the plant model and optimization cost. 

Online Feedback Optimization (OFO) has recently garnered interest as a solution to this challenge. 
By directly interconnecting the optimization process with the true plant, OFO enables continual adaptation and exploits the robustness of feedback control \cite{robust_feedback_controllers, bolognanis_paper, mattia_stability, dallanese}. 
For this reason, OFO has found application in several engineering domains, most prominently power systems (\cite{simpson2020stability,chen2020distributed, Tang2017, Emiliano2016}), but also communication networks \cite{wang2011control}, smart building automation \cite{Belgioiosogiuseppe2021}, traffic
control \cite{bianchin2021time} and process control \cite{ZAGOROWSKA2023119}. 

In its most basic form, OFO aims at solving the following optimization problem online:
\begin{subequations} \label{general_problem}
\begin{align}
    \min_{u\in \R^m, y\in\R^p} \; & \phi_u(u) + \phi_y(y) \label{general_problem_objective}\\
    \text{s.t.} \; & C_u(u) \leq 0 \label{general_problem_input_constr}\\
       & C_y(y) \leq 0 \label{general_problem_output_constr}\\
       &  y =h(u).\label{general_problem_ss_constraint}
\end{align}
\end{subequations}
Here, $u\in \R^m$ and $y\in\R^p$ are the input and output of a physical plant. We model the plant through its input-output map $h: \R^m \rightarrow \R^p$. This map could approximate a stable dynamical system with fast-decaying dynamics, such that the system output quickly converges to $h(u)$ when a constant input $u$ is applied. The map $h$ does not need to be known; instead, OFO relies on online measurements of the output $y$ and on the knowledge (or online estimate \cite{he2022, picallo2022adaptive, chan2025robust}) of the input-output sensitivity $\nabla_u h$. 

In this work, we investigate the use of OFO in a market framework in which control actions are taken by multiple autonomous and self-interested actors, rather than by a single centralized decision-maker. 
In this setup, a market operator aims to determine incentives (rewards or payments to be charged) such that the self-interested actors collectively steer the plant toward a socially optimal outcome.

Implementing OFO in such an online market presents some challenges. 
One major challenge is enforcing the output constraints \eqref{general_problem_output_constr}.
While methods based on Lagrangian duality allow decentralization across the actors \cite{Bernstein2019}, they often suffer from slow convergence and significant constraint violations during transients \cite{DirzaSkogestad}. 
Another issue is that the market operator does not know the actors' costs and constraints, nor can it directly control their actions---instead, actors greedily react to incentives to maximize their profit.
These challenges were considered in \cite{incentive_ofo} for a prosumer incentivization problem. 
However, the bilevel-optimization formulation in \cite{incentive_ofo} differs from the social welfare optimization goal in this paper.  

\smallskip 
\emph{Contributions}:
We present PRIME (PRoximal Iterative MarkEts), a new OFO approach  to coordinate self-interested market participants (``input actors" and ``output actors") attached to a physical plant's inputs and outputs (see  Figure~\ref{fig:actors}).
The method builds on a primal-dual Lagrangian formulation, enhanced with proximal-point iterations, to improve convergence speed and enable non-smooth objective functions.

We introduce two different variants of PRIME.
PRIME-Y constructs the Lagrangian by dualizing \eqref{general_problem_output_constr}, and allows for input actors only. 
 PRIME-H instead constructs the Lagrangian by dualizing the input-output constraints in \eqref{general_problem_ss_constraint}, enabling both input and output actors, and thus a higher degree of decentralization.

PRIME requires the market operator to know only the plant's input-output sensitivity $\nabla_u h$, but not the actors' private costs or constraints. 
Notably, the strictly convex incentive functions constructed by the market operator naturally compel each actor to respond by optimizing their (private) cost. This process  implicitly reveals actors' preferences and constraints without requiring central modeling, which is prone to model mismatch. 

Via simulations on an Optimal Power Flow (OPF) problem, we show that PRIME matches the performance of standard centralized methods in terms of constraint resolution and convergence speed, while benefiting from the distributability of a market-based implementation.


\begin{subfigures}\label{fig:actors}
\begin{figure}
    \centerline{\includegraphics[width=0.5\textwidth]{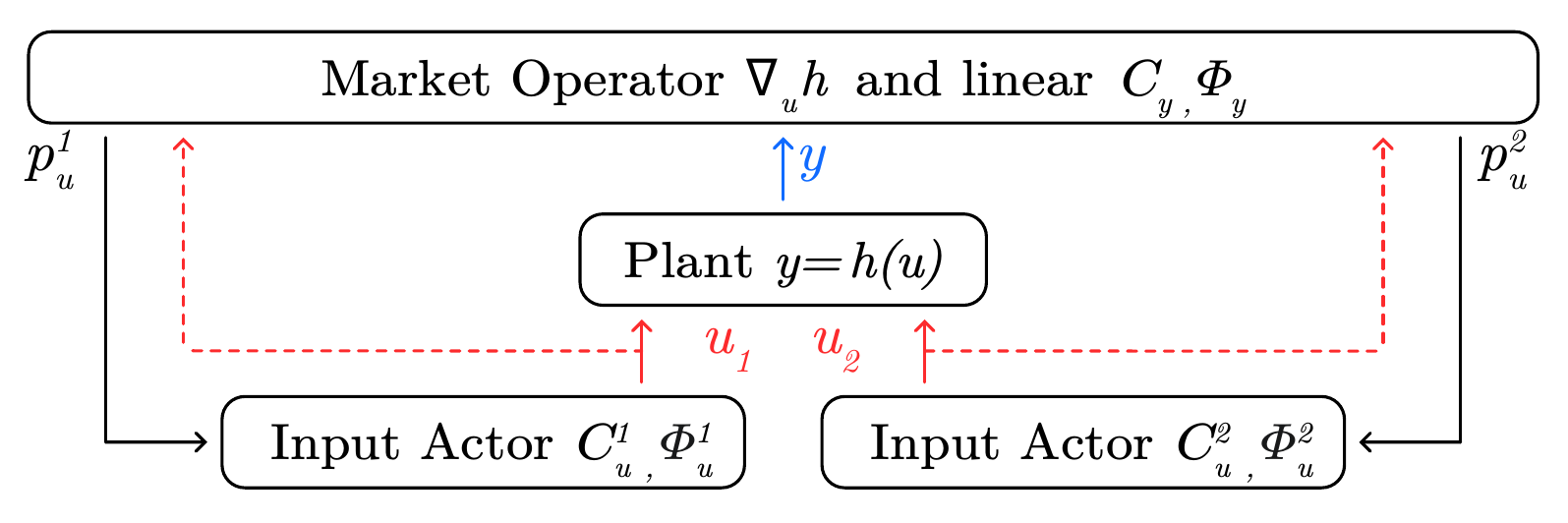}}
    \caption{Market operator coordinating two input actors with PRIME-Y. 
    PRIME-Y does not allow for output actors.}
    \label{fig_prime_y}
\end{figure}
\begin{figure}
    \centerline{\includegraphics[width=0.5\textwidth]{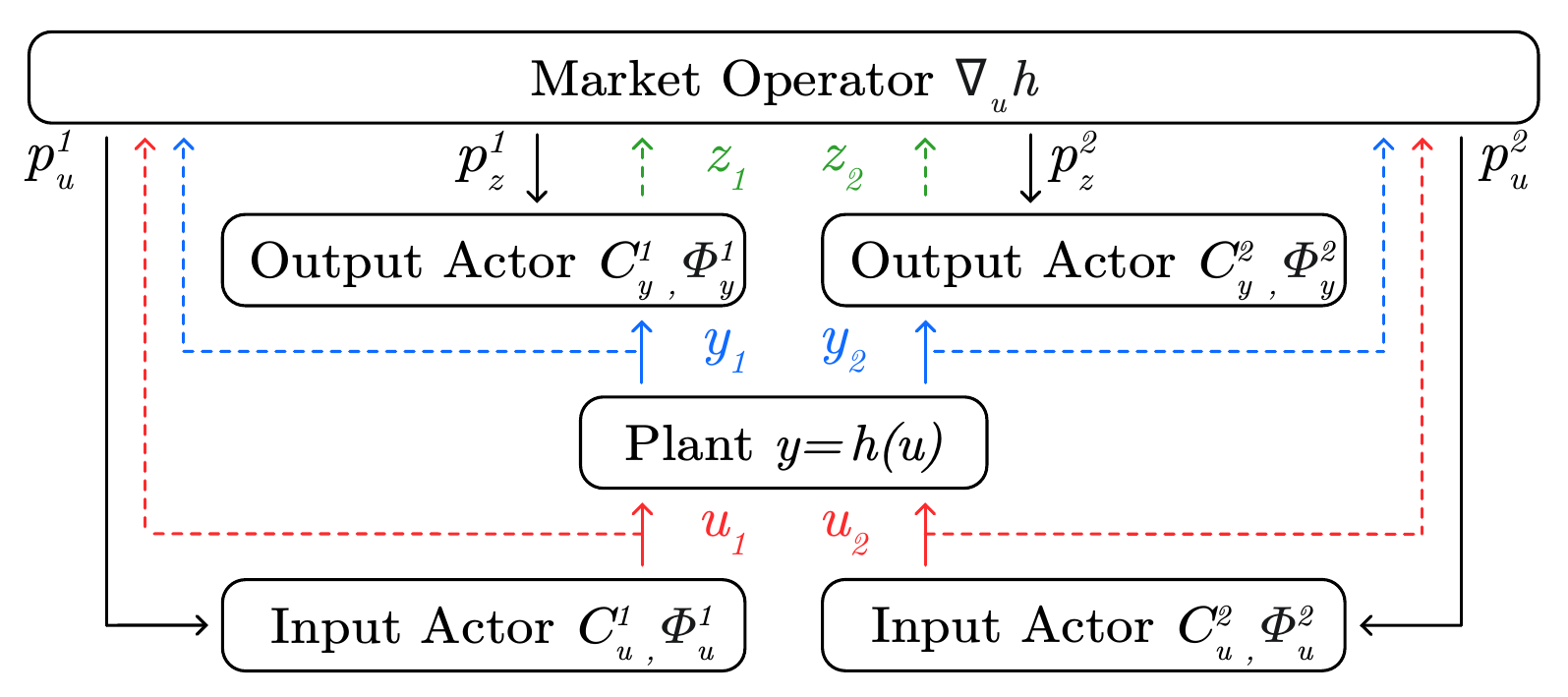}}
    \caption{Market operator coordinating two input actors and two output actors with PRIME-H.}
    \label{fig_prime_h}
\end{figure}
\end{subfigures}

\section{Market model} \label{market}
In this work, we focus on the following instance of \eqref{general_problem}: 
\begin{subequations} \label{general_problem_actor}
\begin{align}
    \min_{u\in \R^m, y\in\R^p} \; & \sum_{j=1}^{n_u} \phi_u^j(u_j) + \sum_{l=1}^{n_y} \phi_y^l(y_l)\\
    \text{s.t.} \; & C_u^j(u_j) \leq 0, \; j \in 1 ,\dots, n_u\\
       & C_y^l(y_l) \leq 0, \; l \in 1 ,\dots, n_y\\
       & y = h(u). \label{eq:plantIO}
\end{align}
\end{subequations}
which represents a market involving $n_u$ input actors and $n_y$ output actors. Here, $\phi_u^j$ and $C_u^j$
are the privately known costs and constraints of input actor $j$. 
Each input actor $j$ wishes to minimize $\phi^j_u$ while satisfying its constraints $C_u^j$, and we denote $u = \begin{bmatrix} u_1^\top & \dots & u_{n_u}^\top \end{bmatrix}^\top$. 
Similarly, $\phi_y^l$ and $C_y^l$ represent the private objective and constraints of the output actor $l$, and  $y = \begin{bmatrix} y_1^\top & \dots & y_{n_y}^\top \end{bmatrix}^\top$.


 Input and output actors are connected via the plant \eqref{eq:plantIO}.
The goal of the market operator is to superpose incentives such that when actors minimize their local cost subject to their private constraints, the plant is steered to a cooperative optimum, i.e., a solution to \eqref{general_problem_actor}.  In particular, note that: 

\begin{enumerate}
    \item incentives, distributed by the market operator, influence the decisions of the actors;
    \item the self-interested actors react to incentives by minimizing their overall cost, given by their private cost function plus the payment due to the market operator.  
\end{enumerate}

This mechanism is illustrated in Figure~\ref{fig_prime_h}.
The input actors directly affect the plant via their decision variables $u_j$.
The output actors, on the other hand, do not have direct control over $y$. However, an output actor $l$ can affect the market dynamics (and indirectly the output $y$) by choosing its decision variable $z_l$, as explained in Section~\ref{sec_our_methods}.

\subsection{Power Systems Example}\label{sec:ps_ex_setup}

We present an OPF problem as an example for the market model in \eqref{general_problem_actor}. 
Consider a distribution grid with $B+1$ buses, including $B$ PQ-buses and one slack bus. The output of the system is given by the vector of the voltages $y = (v_b,\theta_b)_{ b \in  \{1,2,\dots, B\} }$, where $v_b$ is the voltage magnitude and $\theta_b$ the voltage phase of bus $b$.

We consider $n_y$ output actors, each of which is responsible for the voltage  in some subset of the network buses. For example,  $y_1 = (v_1,\theta_1,v_2, \theta_2)$, means that Output Actor 1 is responsible for the voltage at buses 1 and 2. Furthermore, $C_y^1(y_1) =(v_1-1.05,v_2-1.05,-v_1+0.95,-v_2+0.95)$, $\phi_y^1 (y_1) = (v_1-1)^2 + (v_2-1)^2$ means that Output Actor 1 constraints the voltage magnitudes between 0.95 and 1.05
per unit, and minimizes the deviations from the nominal voltage of 1. Output Actor 1 and the other output actors pay the market operator to keep their voltages at the desired levels.

On the distribution grid, there are also $n_u$ input actors/prosumers that adjust their  power injections and pay/are paid by the market operator to do so. Each prosumer $j$ has a power injection $u_j =(P_j,Q_j)$ at a certain bus, where $P_j$ and $Q_j$ are the active and reactive power injection, respectively. 
The input of the system is the vector of power injections, $u = (u_j)_{j \in \{ 1,2,\dots,n_u \}}$. 

The power injections determine the voltages through the power flow  $y = h(u)$. The setup described thus far corresponds to the scheme in Figure~\ref{fig_prime_h}.

It is also possible that there is only one output actor responsible for the voltage magnitudes at all  the buses. In this case, the output actor can coincide with the market operator and can be considered a Distribution System Operator (DSO). This setup corresponds to the scheme in Figure~\ref{fig_prime_y}.

\section{Review of OFO approaches} 
\label{sec_review}

When implementing OFO, the input constraints \eqref{general_problem_input_constr} can easily be satisfied by clipping the inputs. Instead, the difficulty lies in the output constraints \eqref{general_problem_output_constr}, which indirectly restrict the inputs through the unknown input-output map $h(u)$.
Replacing all occurrences of $y$ in \eqref{general_problem} with $h(u)$ leads to the following problem formulation in $u$:
\begin{subequations} \label{primal_problem}
\begin{align} \label{primal_problem_costAA}
    \min_{u\in \R^m} \; & \phi_u(u) + \phi_y(h(u))\\
    \text{s.t.} \; & C_u(u) \leq 0 \label{primal_problem_cu}\\
                   & C_y(h(u)) \leq 0.\label{primal_problem_cy}
\end{align}
\end{subequations}
We review how OFO can be used to solve \eqref{primal_problem} with either a linearized primal formulation (Section \ref{ofo:prim}) or a Lagrange dual formulation (Section \ref{ofo:dualy}).

\subsection{Projected Primal Gradient OFO} \label{ofo:prim}
The linearized primal approach replaces $h$ in \eqref{primal_problem_cy} with $\tilde h^k$,
\begin{align}\label{eq:htilde}
\Tilde{h}^k(u) := y^k + \nabla_u h(u^k)^\top (u - u^k),
\end{align}
which is the linearization of $h$ at the input/output sample obtained from the system at the last timestep, $u^k,y^k$.
The linearized constraint \eqref{primal_problem_cy} is then
\begin{align}\label{eq:linearizedcy}
C_y(y^k) + \nabla_y C_y(y^k)^\top \nabla_u h(u^k) (u  - u^k) \leq 0.
\end{align}
Let $\proj_{\mathcal{U}_k}(u)$ be the projection  onto $\mathcal{U}_k$, the feasible set for $u$ determined by \eqref{primal_problem_cu} and \eqref{eq:linearizedcy}.
Algorithm~\ref{alg:primal} describes Projected Primal Gradient OFO \cite{nonconvex_optimization, phd_hauswirth}, 
obtained using  feedback from the plant, i.e., replacing $h(u^k)$ in \eqref{primal_problem_costAA} with the measured $y^k$.

\begin{algorithm}[htbp]
    \caption{Projected Primal Gradient OFO}\label{alg:primal}
\begin{algorithmic}
    \STATE $\alpha \in \mathbb{R}_{>0}$ \COMMENT{primal learning rate}
    \STATE $u^0 \gets u^\text{initial}$
    \FORALL{$k=0,1,\dots$}
        \STATE $y^k \gets h(u^k)$ \COMMENT{apply u to the system and sample y}
        \STATE $\hat{u}^k \gets u^k - \alpha \, (\nabla_u\phi_u(u^k) + \nabla_u h(u^k)^\top \nabla_y \phi_y(y^k))$
        \STATE $u^{k+1} \gets \proj_{\mathcal{U}_k}(\hat{u}^k)$
    \ENDFOR 
    \end{algorithmic}
\end{algorithm}

If problem \eqref{primal_problem} is convex, and in the absence of measurement noise, one can obtain global convergence guarantees for Algorithm~\ref{alg:primal} to a solution of \eqref{primal_problem} \cite{nonconvex_optimization}.
However, if $h$ is nonlinear or in the presence of measurement noise, the set $\mathcal U_k$ can be empty, breaking recursive feasibility; for an example, see the Appendix in the extended draft \cite{behr2025primefastprimaldualfeedback}.
Furthermore, Algorithm~\ref{alg:primal} is a \emph{centralized} method that cannot be implemented as a market.

\subsection{Primal-Dual OFO: dualizing the output constraint} \label{ofo:dualy}
Instead of using a projection, the output constraint in \eqref{primal_problem_cy} may be dualized by forming the Lagrangian \eqref{lagrange_dual_dualy}, where $\Psi_{C_u}(u)$ is the indicator function of the input constraints, which is zero when $C_u(u)\leq0$ and $\infty$ otherwise:
\begin{equation}\label{lagrange_dual_dualy}
\begin{split}
    \mathcal{Y}(u, \lambda_y) &= \phi_u(u) + \phi_y(h(u)) 
    + C_y(h(u))^\top \lambda_y + \Psi_{C_u}(u).
\end{split}
\end{equation}
Given this formulation, one can seek a solution to \eqref{primal_problem} by performing gradient descent/ascent, by minimizing the Lagrangian for the primal variable $u$ with learning rate $\alpha$ and maximizing for the dual variable $\lambda_y$ with learning rate $\rho$, and introducing feedback from the system, as outlined in Algorithm \ref{alg:dualy}.
We define $\proj_{C_u}(u)$ as the projection of $u$ onto the constraint $C_u(u) \leq 0$, which differs from $\proj_{\mathcal{U}_k}(u)$ as it does not include \eqref{eq:linearizedcy}.
Under regularity conditions, this algorithm is guaranteed to converge to the optimal solution on convex problems \cite{robust_feedback_controllers}.

\begin{algorithm}[htbp]
    \caption{Primal-Dual OFO, dualizing output constraint} \label{alg:dualy}
\begin{algorithmic}
    \STATE $\alpha \in \mathbb{R}_{>0}$ \COMMENT{primal learning rate}
    \STATE $\rho \in \mathbb{R}_{>0}$ \COMMENT{dual learning rate}
    \STATE $u^0\gets u^\text{initial}$, $\lambda_y^0 \gets 0$
    \FORALL{$k=0,1,\dots$}
        \STATE $y^k \gets h(u^k)$ \COMMENT{apply u to the system and sample y}
        \STATE $\lambda_y^{k+1} \gets \proj_{\mathbb{R}_{\geq0}}(\lambda_y^k + \rho C_y(y^k))$
        \STATE $\nabla_u \mathcal{Y}(u^k, \lambda_y^{k+1}) \gets \nabla_u\phi_u(u^k)$
        \STATE $\hspace{4.5mm}+\;\nabla_u h(u^k)^\top ( \nabla_y \phi_y(y^k) + \nabla_y C_y(y^k)^\top \lambda_y^{k+1})$
        \STATE $u^{k+1} \gets \proj_{C_u}(u^k - \alpha \nabla_u \mathcal{Y}(u^k, \lambda_y^{k+1}))$
    \ENDFOR 
\end{algorithmic}
\end{algorithm}

As standard in Lagrange dual methods, Algorithm~\ref{alg:dualy} offers a price interpretation through its dual multipliers $\lambda_y$, which ensure satisfaction of \eqref{general_problem_output_constr}.
However, Algorithm~\ref{alg:dualy} is still ill-suited for a market of \emph{rational} actors (i.e., actors aiming at maximizing their profit), who would update their variables $u$ as their best response to $\lambda_y$, and not take gradient steps as asserted by Algorithm~\ref{alg:dualy}.
Furthermore, primal-dual gradient methods are known to be slow, as demonstrated in Figure~\ref{fig_unicorn}.

\section{Our Proposed Methods} \label{sec_our_methods}

\subsection{PRIME-Y OFO: dualizing the output constraint} \label{ofo:primey}
To address the challenges discussed for Algorithm~\ref{alg:dualy}, we employ proximal minimization \cite{prox_algs} instead of gradient descent to update the variable $u$. In particular, we would like to perform $u^{k+1} = \operatorname{prox}_{\mathcal Y(\cdot,\lambda_y^{k+1})/\gamma_u} (u^k)$, defined by 
$$u^{k+1} = \argmin_u \ \mathcal{Y}(u,\lambda_y^{k+1})+\frac{\gamma_u}{2} \|u-u^k\|^2.$$
However, this update cannot be performed, as $h$ is unknown. Instead, we approximate the Lagrangian \eqref{lagrange_dual_dualy} by $\tilde{\mathcal{Y}}^k(u) = \Tilde{\mathcal{Y}}(u,u^k, y^k,\lambda_y^{k+1})$, where $h$ is replaced by its linearization \eqref{eq:htilde} at timestep $k$, 
\begin{align}
    \begin{split}
        \mathcal{\tilde Y}^k(u) &= \phi_u(u) + \phi_y(\tilde h^k(u))\\
        & \qquad + C_y(\tilde h^k(u))^\top \lambda_y^{k+1} + \Psi_{C_u}(u)
    \end{split} \label{eq:Ltilde1} \\
    \begin{split}
        u^{k+1} &= \prox_{\Tilde{\mathcal{Y}}^k/\gamma_u}(u^k)\\
        & = \argmin_{ u} \ \Tilde{\mathcal{Y}}^{k}(u) + \frac{\gamma_u}{2}\|u - u^k\|_2^2.
    \end{split}\label{prox_dualy}
\end{align}
The second summand of \eqref{prox_dualy} can be interpreted as a damping term, encouraging the iterates to remain close to the current $u^k$, where the linearization approximation holds. Algorithm~\ref{alg:prox_dualy} summarizes the resulting OFO method, PRIME-Y. 

\begin{algorithm}[htbp]
    \caption{PRIME-Y} \label{alg:prox_dualy}
\begin{algorithmic}
    \STATE $\gamma_u \in \mathbb{R}_{\geq0}$ \COMMENT{proximal deviation cost}
    \STATE $\rho \in \mathbb{R}_{>0}$ \COMMENT{dual learning rate}
    \STATE $u^0\gets u^\text{initial}$, $\lambda_y^0 \gets 0$
    \FORALL{$k=0,1,\dots$}
        \STATE $y^k \gets h(u^k)$ \COMMENT{apply u to the system and sample y}
        \STATE $\lambda_y^{k+1} \gets \proj_{\mathbb{R}_{\geq0}}(\lambda_y^k + \rho\, C_y(y^k))$
        \STATE $u^{k+1} \gets \prox_{\Tilde{\mathcal{Y}}^k/\gamma_u}(u^k)$ \COMMENT{$\Tilde{\mathcal{Y}}^k$ from Equation \eqref{eq:Ltilde1}}
    \ENDFOR
\end{algorithmic}
\end{algorithm}

For example, in the case of quadratic input cost $\phi_u(u) = u^\top Q_u u + c_u^\top u$,
{linear} output cost $\phi_y(y) = c_y^\top y$, and {linear} output constraints $C_y(y) = A_y y - b_y\leq0$, the proximal update is solved by the following optimization:
\begin{subequations} \label{qp_dual_y_u}
\begin{align}
    Q &= Q_u + \frac{\gamma_u}{2} I\\
    \begin{split}
    c &= c_u - \gamma_u u^k + \nabla_u h(u^k)^\top (A_y^\top \lambda_y^{k+1}  + c_y)
    \end{split}\label{eq:linearcostconstraints}\\
    u^{k+1} &= \argmin_{u \in C_u} u^\top Q u + c^\top u.
\end{align}
\end{subequations}

\subsubsection{Market Interpretation of PRIME-Y}
As per Section \ref{market}, a market algorithm should superpose incentives on the private actor costs so that the aggregate behavior converges to a cooperative optimum. 
Specifically, we would like to choose a payment $p_u^{j,k} (u_j)$ from each input actor $j$ at timestep $k$ such that the combined best responses of the input actors $j$,
\begin{align}\label{eq:BR}
     u_j^{k+1} = \argmin_{u_j \in C_u^j} \ \hat \phi_u^{j,k}  (u_j) := \phi_u^j (u_j)+ p_u^{j,k} (u_j),
\end{align}
mimics the update in \eqref{prox_dualy}. 
If the payment is positive, the actor pays the market operator, and if the payment is negative, the market operator pays the actor.

Unfortunately, it is not always possible for PRIME-Y's  $n_u$ best responses \eqref{eq:BR} to match \eqref{prox_dualy}, as the update in \eqref{prox_dualy} is not \emph{generally} separable in $u$, because $\phi_y(\tilde h^k(u))$ and $C_y(\tilde h^k(u))$ couple the costs between inputs. 
A notable exception is the case in which the output cost and constraints are \emph{linear} as in \eqref{eq:linearcostconstraints}; in this scenario, the incentive payment
\begin{equation}
\begin{split}\label{price_prime_y}
p_u^{j,k}(u_j) &= \frac{\gamma_u}{2} (u_j-u_j^k)^\top(u_j-u_j^k)\\
& \qquad + u_j^\top e_j ^\top  \nabla_u h(u^k)^\top( A_y^\top \lambda_y^{k+1} + c_y),
\end{split}
\end{equation}
where $e_j$ is the matrix such that $u_j = e_j^\top u$, achieves the goal of mimicking \eqref{prox_dualy} with the $n_u$ best responses \eqref{eq:BR}. 
With the additional assertion of quadratic input costs, the aggregate best response coincides with the update \eqref{qp_dual_y_u}.

Computing the PRIME-Y incentive \eqref{price_prime_y} requires the knowledge of the output costs ($c_y$) and constraints ($A_y$). 
Since the market operator requires this information, PRIME-Y does not allow for autonomous output actors.
Instead, the market operator is responsible for the output cost and constraints. 
Another way of seeing this is that the market operator coincides with the only output actor.

\subsection{PRIME-H OFO: dualizing the input-output constraint} \label{ofo:primeh}

Algorithms \ref{alg:primal}, \ref{alg:dualy}, \ref{alg:prox_dualy} exploit the formulation \eqref{primal_problem}, which is obtained by replacing $y$ in \eqref{general_problem} with the input-output map $h(u)$. 
While this substitution eliminates $y$, it couples the inputs and outputs, which prevents output actor distribution and, in the case of nonlinear output constraints or costs, also prevents input actor distribution.
Thus, exploiting the formulation in \eqref{primal_problem}, as done in PRIME-Y, may not be well-suited for market implementations.

Rather than substituting $h(u)$ for $y$, we propose to dualize the input-output map constraint \eqref{eq:plantIO}.
To do so, we introduce an auxiliary variable $z$, which is a proxy for the output $y$, and we define the Lagrangian
\begin{equation}\label{dualh_problem}
\begin{split}
\mathcal{H}(u, z, \nu_h) &= \phi_u(u) + \phi_y(z) + (h(u) - z)^\top \nu_h\\ & \qquad + \Psi_{C_u}(u) + \Psi_{C_y}(z).
\end{split}
\end{equation}
The Lagrangian has two primal variable vectors, $u$ and $z$, that correspond to the decisions of the input and output actors, respectively; and a dual variable $\nu_h$, related to the constraint $z = h(u)$. The variable $z$ can be understood as a target for the physical output $y$. Due to the presence of the indicator function  $\Psi_{C_y}(z)$, the ``output target'' $z$ will always satisfy the output constraints $C_y(z) \leq 0$.

%

To seek a saddle-point of the Lagrangian $\mathcal H$, we would again like to leverage a proximal iteration \cite{prox_algs}.
However, as $h$ is  unknown, $\mathcal H$  is first approximated by $\Tilde{\mathcal{H}}^k(u, z) = \Tilde{\mathcal{H}}(u, z, u^k, y^k, \nu_h^{k+1})$ where $h$ is replaced by its linearization $\Tilde{h}^k$ \eqref{eq:htilde} at timestep $k$:
\begin{equation}\label{eq:HTILDE}
\begin{split}
\Tilde{\mathcal{H}}^k(u, z) &= \phi_u(u) + \phi_y(z) + (\Tilde{h}^k(u) - z)^\top \nu_h^{k+1}\\ & \qquad + \Psi_{C_u}(u) + \Psi_{C_y}(z).
\end{split}
\end{equation}
As with \eqref{prox_dualy}, we use the proximal operator, defined as 
\begin{align*}
    \prox_{g/\gamma}(u^k) = \argmin_{u} \ g(u) +  \frac{\gamma}{2}\|u - u^k\|_2^2.
\end{align*}
To improve convergence, we augment the $z$ update with a quadratic damping term,
\begin{align}
f^k(z) &= \Tilde{\mathcal{H}}^k(u^k, z) +  \frac{\rho}{2} \| y^k - z \|_2^2, \label{augmented_lagrangian}
\end{align}
that pushes $z$ towards the previously measured output $y^k$.

The resulting feedback controller, PRIME-H, is shown in Algorithm~\ref{alg:prox_dualh}.
Note that the update of $\nu_h$ can be seen as an integrator which enforces $z=y$ in the limit. 
Due to the proximal updates on both $u$ and $z$, PRIME-H allows for non-smooth costs for both the input and output. Furthermore, as shown in the following, PRIME-H can be distributed over \emph{both} input and output actors, for arbitrary constraints and cost functions as in \eqref{general_problem_actor}. 

\begin{algorithm}[htbp]
    \caption{PRIME-H} \label{alg:prox_dualh}
\begin{algorithmic}
    \STATE $\gamma_u \in \mathbb{R}_{\geq0}$ \COMMENT{proximal deviation cost in u}
    \STATE $\gamma_z \in \mathbb{R}_{\geq0}$ \COMMENT{proximal deviation cost in z}
    \STATE $\rho \in \mathbb{R}_{>0}$ \COMMENT{dual learning rate}
    \STATE $u^0 \gets u^\text{initial}$, $z^0 \gets proj_{C_y}(h(u^0)), \nu_h^0 \gets 0$
    \FORALL{$k=0,1,\dots$}
        \STATE $y^k \gets h(u^k)$ \COMMENT{apply u to the system and sample y}
        \STATE $\nu_h^{k+1} \gets \nu_h^k + \rho \, (y^k - z^k)$
        \STATE $z^{k+1} \gets \prox_{f^k/\gamma_z}(z^k)$ \COMMENT{$f^k$ from   Equation \eqref{augmented_lagrangian} }
        \STATE $u^{k+1} \gets \prox_{\Tilde{\mathcal{H}}^k(u, z^k)/\gamma_u}(u^k)$ \COMMENT{$\tilde {\mathcal H}^k$ from Equation \eqref{eq:HTILDE}}
    \ENDFOR 
\end{algorithmic}
\end{algorithm}

\subsubsection{Market Interpretation of PRIME-H}
The market operator provides the following incentive to each input actor $j$, \vspace{-1em}
\begin{subequations}\label{penal_dual_h}
\begin{align}
\begin{split}
    p_u^{j,k}(u_j) &= \frac{\gamma_u}{2} (u_j-u_j^k)^\top (u_j-u_j^k)\\
    & \qquad + u_j ^\top e_j^\top  \nabla_u h(u^k)^\top \nu_h^{k+1}, \label{penal_dual_h_u}
\end{split}
\end{align}
and the following incentive to each output actor $l$,
\begin{align}
\begin{split}
    p_z^{l,k}(z_l) &= \frac{\rho + \gamma_z}{2} (z_l-z
    _l^k)^\top (z_l-z_l^k) \\ 
    & \qquad  - z_l^\top  g_l^\top (\nu_h^{k+1} + \rho (y^k - z^k)), \label{penal_dual_h_z}
    \end{split}
\end{align}
\end{subequations}
where $g_l$ is the matrix such that $y_l = g_l^\top y$.

It is easy to check that the best response (as in \eqref{eq:BR}) of the input and output actors coincides with the update of $z$ and $u$ in Algorithm~\ref{alg:prox_dualh}. 

\emph{Role of the output actors}: the output actors determine $z$ and thus the target for $y$. While the output actors do not directly change the system inputs, their decisions for $z$ affect $\nu_h$ and the input actor incentives ($p_u^{j,k}$), and thus coerce the input actors to adjust the inputs $u$ so that the output $y$ eventually matches $z$. 

Moreover, note that the market operator does not need any information on the input/output cost or constraints to compute $\nu_h^{k+1}$ and the incentives $p_u^{j,k}$ and $p_z^{l,k}$. Therefore, PRIME-H admits a market implementation and can be distributed \emph{over both input and output actors}, preserving the privacy of the cost functions and constraints of both.
This differs from PRIME-Y, which does not allow for independent output actors and can be distributed over input actors only in the case of linear output costs and constraints.
Table \ref{deg-of-cent-table} summarizes the market interpretation distinctions.

\begin{table}[H]
\centering
\caption{Degree of Decentralization}
\label{deg-of-cent-table}
\begin{tblr}{
  cells = {c},
  vline{2} = {-}{},
  hline{2} = {-}{},
}
 & Algorithm \ref{alg:primal} & PRIME-Y & PRIME-H\\
Input Actor Market & \ding{55} & (\ding{51}$^*$) & \ding{51}\\
Output Actor Market & \ding{55} & \ding{55} & \ding{51}
\end{tblr}
\textsuperscript{$*$ In case of linear output cost and constraints.}
\end{table}

\section{Comparison of Projected Primal Gradient \& PRIME Methods} \label{sec_comparison}

\begin{figure}[htbp]
\centerline{\includegraphics[width=0.5\textwidth]{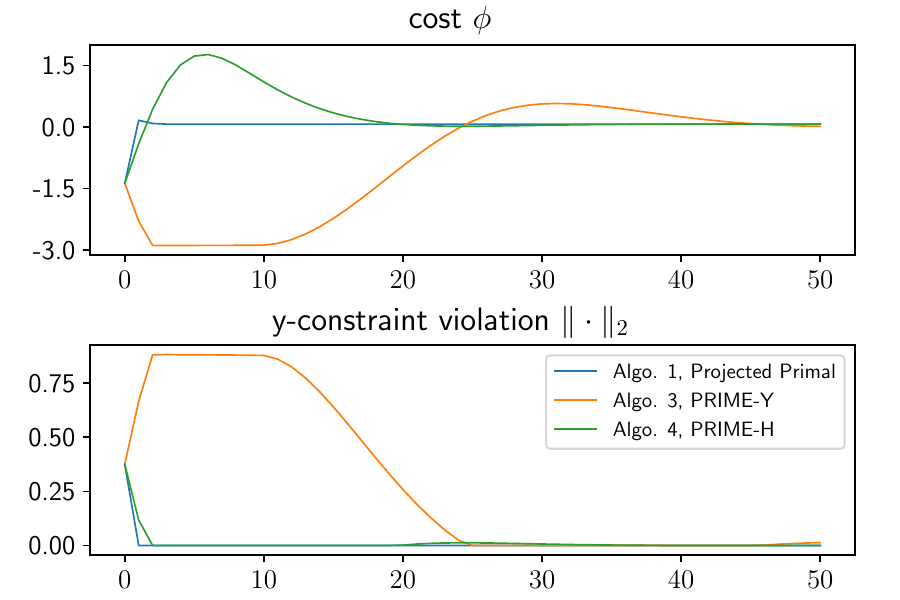}}
\caption{Toy example with non-zero output cost and no measurement noise.}
\label{fig_nonconvex_toy}
\end{figure}

We now illustrate the difference between PRIME-Y and PRIME-H in a small numerical example in Figure~\ref{fig_nonconvex_toy}. 
For the map $y=h(u)=u_2^3+u_1-u_2+0.5$ we minimize $\phi_u(u) +\phi_y(y)  = u_1^2 - 0.5 u_1 + u_2^2 - 0.5 u_2+ 5y$ on 
$\mathcal{U}=[-1,1]^2$ and $\mathcal{Y}=[0,1]$, and initialize at $u=[-0.5, 0.5]$.
PRIME-Y and PRIME-H differ in the treatment of output costs $\phi_y$.
In particular, PRIME-Y heads for the unconstrained optimum along the composition $\nabla_u \phi_y(h(u))$, initially achieving a lower cost while increasing the output constraint violation, and only later, when the dual variables are built, obeying the output constraints.
Instead, PRIME-H targets ``legal" improvements to $y$ via $z$, initially focusing on resolving violations, and only later, when the dual variables are built, optimizing the cost over the input-output map $h$.
Finally, Figure~\ref{fig_nonconvex_toy} also shows the behavior of Algorithm~\ref{alg:primal}. While Algorithm~\ref{alg:primal} outperforms the PRIME methods in this example, it is essential to note that Algorithm~\ref{alg:primal} is not a distributed market algorithm and requires central access to the exact cost and constraint functions. We will also show in the next section that  Algorithm~\ref{alg:primal} is less robust to measurement noise.

\section{Application to Power Grid Operation} \label{application}

\subsection{Power Network Model}
Consider a grid with $B+1$ buses, as described in Section~\ref{sec:ps_ex_setup}, with  $B$ PQ-buses and one slack bus.
The net real power injections $P \in \mathbb{R}^B$ and reactive power injections $Q \in \mathbb{R}^B$ from the prosumers are stacked to create the input vector $u$, and the voltage magnitudes and phase angles at all non-slack buses are stacked to form the output vector $y \in \mathbb{R}^{2B}$.
For simplicity, every nodal net power injection is a distinct input actor, and there is a single DSO output actor.



Following the approach outlined in \cite[Ch. 6]{moffat2022dissertation} and \cite{bolognanis_paper}, we obtain the voltage-power sensitivity 
$\Gamma \in \mathbb{R}^{2B \times 2B}$ at the operating point $(u^*,y^*)$: 
\begin{align*}
    \Gamma   \delta y = \delta  u,
\end{align*}
where $\delta y$ is a small change in the output $y$ coresponding to $\delta  u$, a small change in the input $u$.
Inverting $\Gamma$, we get the desired power-voltage sensitivity $\nabla_uh$: 
\begin{align*}
 \delta y = \nabla_u h \; \delta u.
\end{align*}
Replacing $\delta y$ with $(y - y^k)$ and $\delta u$ with $(u - u^k)$, we get $\Tilde{h}^k(u)$, defined in \eqref{eq:htilde}, which linearizes $h$ at timestep $k$. 

\begin{figure}
    \centering
    \begin{subfigure}{\linewidth}
        \centering
        \includegraphics[width=\linewidth]{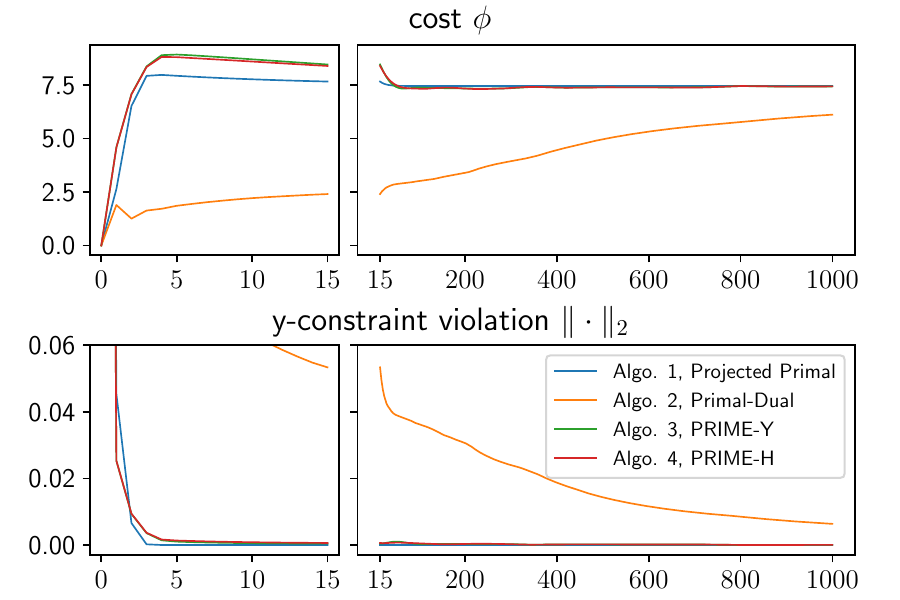}
        \caption{No measurement noise}
        \label{fig_unicorn_nonoise}
    \end{subfigure}
    \begin{subfigure}{\linewidth}
        \centering
        \includegraphics[width=\linewidth]{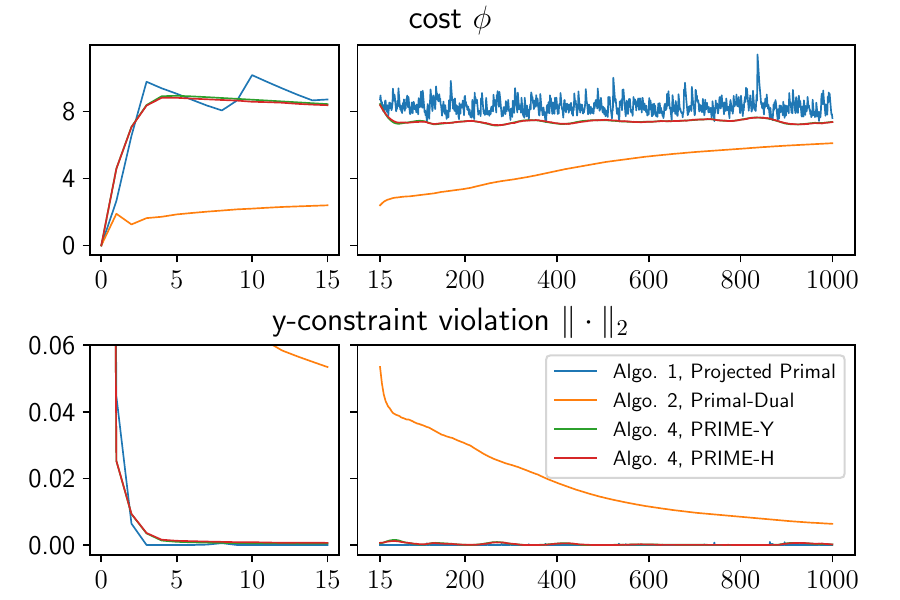}
        \caption{Additive Gaussian voltage measurement noise with a standard deviation of 1.5\% of the allowable voltage range ($0.15\%$ per unit).}
        \label{fig_unicorn_noise}
    \end{subfigure}
    \caption{Coldstart turn-on performance on the UNICORN 56-bus \cite{unicorn} test case with two active-\& reactive power controllable generators and 50 uncontrollable loads. The bus voltages are constrained from 0.95 to 1.05 per unit.}
    \label{fig_unicorn}
\end{figure}

\subsection{Performance}
Figure~\ref{fig_unicorn} displays the performance of the methods discussed on the UNICORN 56-bus \cite{unicorn} test case. 
The input actors are divided between prosumers, with cost of production  $\phi_u^j(u_j) = 0.1 P_j + 0.1 P_j^2 + 0.1 Q_j^2$ and production  limited to box constraints $P_j\in[0, 12.5]$ and $Q_j\in[-2, 2]$, and consumers, treated as  fixed loads, e.g. $P_j=[-0.4]$ and $Q_j=[-0.2]$.
Voltage magnitudes are constrained to $[0.95, 1.05]^B$ per unit on all buses. 

For all methods, the dual variables are initialized at 0.
This cold start is challenging, as the dual variables must be built up before constraint satisfaction.
For this reason, the primal-dual method in Algorithm~\ref{alg:dualy} is slow in resolving constraint violations and is outperformed by the Projected Primal method.
Notably, the Projected Primal method requires a cost and constraint oracle (i.e., it is centralized).
Remarkably, PRIME-Y and PRIME-H operate in a market fashion without this oracle and still achieve comparable resolution speeds.

In Figure \ref{fig_unicorn_noise}, we introduce noise into the voltage measurements. The Projected Primal algorithm is sensitive to noise, while the primal-dual methods are not.
The Projected Primal method performs particularly poorly when operating close to output constraints (which is common for optimal points), as measurement noise frequently causes constraints to appear violated, leading the following input to be projected away from the optimum. This aggressive reaction to the measurement noise leads to the Projected Primal method's ``jittery'' behavior in Figure \ref{fig_unicorn_noise}.
PRIME-Y and PRIME-H instead deliver fast constraint violation resolution while rejecting the effect of the measurement noise.

\section{Conclusion} \label{sec_conclusion}
This paper introduced PRIME, a new feedback optimization approach based on proximal iterations. PRIME achieves rapid constraint violation resolution and robustness to measurement noise. 
Most importantly, PRIME enables the design of optimal incentives in markets with self-interested actors without requiring access to their private costs or constraints.

Future work should address the possible strategic behavior of market actors.
This includes examining the risk of market dynamics manipulation and devising mechanisms to make PRIME truth-revealing and collusion-proof.


\bibliographystyle{IEEEtran}
\bibliography{bibliography}

\newpage
\section{Appendix}
\subsection{Projected Primal Gradient OFO Infeasibility}\label{primal_algo_infeasibility}
The Projected Primal Gradient method (Algorithm~\ref{alg:primal}) works by projecting $u^k$ onto the set  $\mathcal{U}^k$, which is the intersection of the input constraints \eqref{primal_problem_cu} and the linearized output constraints \eqref{eq:linearizedcy}. 
This intersection can be empty, resulting in the method returning no solution after initialization, a change in the constraints, or during the progression of gradient steps.

Figure~\ref{primal_initial_infeas} demonstrates how a nonlinear $h(u)$ can lead to Algorithm~\ref{alg:primal} being infeasible at initialization or after a large gradient step.
Figure~\ref{primal_meas_infeas} demonstrates how measurement noise can result in infeasibility as the linearization is no longer made around the true operating point $y^k$.
Note that in both cases, feasible inputs to the true nonlinearized formulation exist, specifically any inputs where the blue line is contained in the dotted perimeter.

\begin{figure}
    \centering
    \begin{subfigure}{\linewidth}
        \centering
        \includegraphics[width=\linewidth]{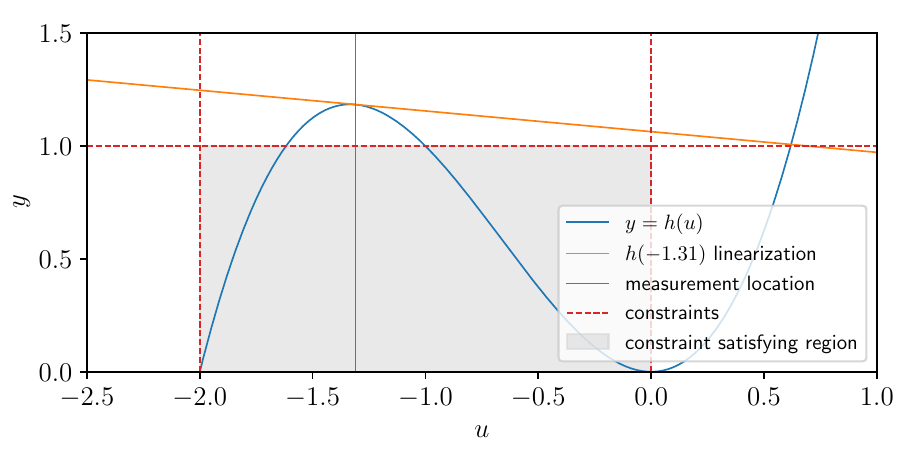}
        \caption{Infeasibility for the constraints $-2 \leq u \leq 0$ and $y \leq 1$ when $u^k=-1.31$ due to the nonlinearity of $h(u)$. The set of allowable next inputs $\mathcal{U}_k  = [-2, 0] \cap [0.7, \infty)$ is empty.}
        \label{primal_initial_infeas}
    \end{subfigure}
    \begin{subfigure}{\linewidth}
        \centering
        \includegraphics[width=\linewidth]{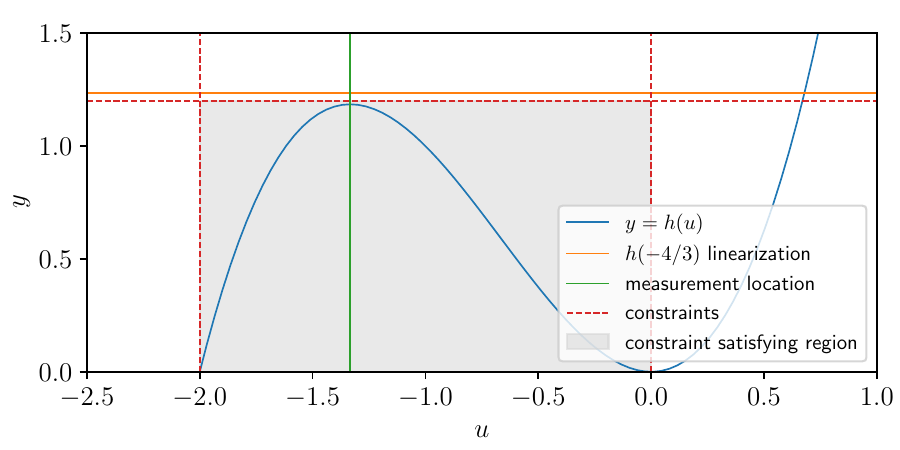}
        \caption{Infeasibility for the constraints $-2 \leq u \leq 0$ and $y \leq 1.2$ when $u^k = -\frac{4}{3}$ due to measurement noise on the output $y$. The noisy measurement $y^k = 1.22$ is not feasible (although the true $y^k = 1.19$ is), and $\nabla_u h (u^k) = 0$ produces an empty $\mathcal{U}_k$.}
        \label{primal_meas_infeas}
    \end{subfigure}
    \caption{Examples of Projected Primal Gradient OFO infeasibility for the input-output map $y = h(u) = 2u^2+u^3$.}
    \label{fig_pimal_infeas}
\end{figure}

\end{document}